\documentclass[conference]{IEEEtran}

\pdfoutput=1

\ifCLASSINFOpdf

\else

\fi

\usepackage[ruled]{algorithm2e}
\usepackage{color}
\usepackage{xcolor}
\usepackage{float}
\usepackage{multirow}
\usepackage{enumerate}
\usepackage{graphicx}
\usepackage{color, colortbl}
\usepackage[font=small]{caption}
\usepackage{subfig}
\usepackage{caption}
\usepackage{graphicx}
\usepackage{amsmath}
\usepackage{framed}
\usepackage{array}
\usepackage{caption}
\usepackage{tcolorbox}
\newcolumntype{P}[1]{>{\centering\arraybackslash}p{#1}}
\definecolor{Gray}{gray}{0.85}
\definecolor{Grays}{gray}{0.95}
\usepackage{adjustbox}
\usepackage{flushend}
\usepackage{varwidth}
\usepackage{framed}
\usepackage{tikz}

\usepackage{tcolorbox}
\usepackage{tcolorbox}
\usepackage{cite}

\usepackage{amsthm}
\usepackage[framemethod=tikz]{mdframed}
\usepackage{lipsum}

\newmdenv[
innertopmargin=0pt,
roundcorner=5pt,
linewidth=1pt,
linecolor=black,
innertopmargin=\baselineskip,
singleextra={
  \node[
    anchor=west,
    xshift=7pt,
    fill=gray!30,
    rounded corners=2pt,
    draw] at (P-|O) {\bfseries Temporal Rules};
},
firstextra={
  \node[
    anchor=west,
    xshift=7pt,
    fill=gray!35,
    rounded corners=2pt,
    draw] at (P-|O) {\bfseries Temporal Rules};
}
]{definition}

\ifCLASSINFOpdf

\else

\fi

\begin{document}
\title{\huge Let's hear it from RETTA: A  Requirements Elicitation Tool for TrAffic management systems  }

%\author{\IEEEauthorblockN{Zahra Shakeri Hossein Abad, Guenther Ruhe}
%\IEEEauthorblockA{Department of Computer Science\\ University of Calgary, Calgary, Canada\\
%Email: \{zshakeri, ruhe\}@ucalgary.ca}
%\and
%\IEEEauthorblockN{Oliver Karras, Kurt Schneider}
%\IEEEauthorblockA{Software Engineering Group\\
%Leibniz UniversitŠt Hannover, Hannover, Germany\\
%Email: \{oliver.karras, kurt.schneider\}@inf.uni-hannover.de}
%
%\IEEEauthorblockN{Parisa Ghazi, Kurt Schneider}
%\IEEEauthorblockA{Software Engineering Group\\
%Leibniz UniversitŠt Hannover, Hannover, Germany\\
%Email: \{oliver.karras, kurt.schneider\}@inf.uni-hannover.de}}

\author{
    \IEEEauthorblockN{Mohammad Noaeen\IEEEauthorrefmark{1}, Zahra Shakeri Hossein Abad\IEEEauthorrefmark{2}, Behrouz Homayoun Far\IEEEauthorrefmark{1}}
    \IEEEauthorblockA{\IEEEauthorrefmark{1}Department of Electrical and Computer Engineering, {\IEEEauthorrefmark{2}Department of Computer Science}} University of Calgary, Calgary, Canada
    \\\{mohammad.noaeen, zshakeri, far\}@ucalgary.ca
}

\maketitle

\begin{abstract}
The area of Traffic Management (TM) is characterized by uncertainty, complexity, and imprecision. The complexity of software systems in the TM domain which contributes to a more challenging Requirements Engineering (RE) job mainly stems from the diversity of stakeholders and complexity of requirements elicitation in this domain. This work brings an interactive solution for exploring functional and non-functional requirements of software-reliant systems in the area of traffic management. We prototyped the RETTA tool which leverages the wisdom of the crowd and combines it with machine learning approaches such as Natural Language Processing and Na\"{i}ve Bayes to help with the requirements elicitation and classification task in the TM domain. This bridges the gap among stakeholders from both areas of software development and transportation engineering. The RETTA prototype is mainly designed for requirements engineers and software developers in the area of TM and can be used on Android-based devices. 
\end{abstract}

\vspace{-2mm}
\section{Introduction}
Developing software-reliant systems for complex domains such as the Traffic Management (TM) domain requires a more challenging RE job. While, in recent decades, a wide range of requirements elicitation techniques has been developed to address complex systems, these techniques usually aim to work in a context-free spectrum and thus are not concerned about the complexity of a particular domain, such as the TM domain. In this domain, we deal with many constraints to elicit requirements, such as the diversity of stakeholders, a clear need for time-centric elicitation techniques, the legal issue for some types of elicitation techniques, and the variability of the transportation demand, knowledge of the road network and the traffic conditions. Moreover, some of the software tools in this domain are critical because they involve people's life (e.g. emergency control systems) and are very sensitive to mistakes. 

To address these issues and following our recent study on exploring transportation engineers' concerns on social Q\&A websites \cite{BIDMA}, we introduce the RETTA Tool, which aims to tackle problems associated with eliciting requirements for TM services such as Emergency Medical Services (EMS), Traffic Signal Timing (TST), and Urban Transportation Planning (UTP) systems. Given that in most of the software tools in the TM domain, the main stakeholders are the crowd participants in a traffic network, who use TM services and at the same time define the requirements of these services, we can rely on them to cater for their needs properly. Thus we apply the {\it crowdsourcing} \cite{Crowd} technique (e.g. social networks mining) as one of the requirements elicitation techniques to prototype the RETTA Tool.  Moreover, we used various quantitative and qualitative data analysis and machine learning techniques to explore TM services requirements from various sources of traffic data, such as real-time camera  and drone data, signal light sensors' data, and historical traffic data.

\vspace{-1mm}
\section{RETTA  Prototype}
\begin{figure}[H]
\centering
\includegraphics[scale=0.05]{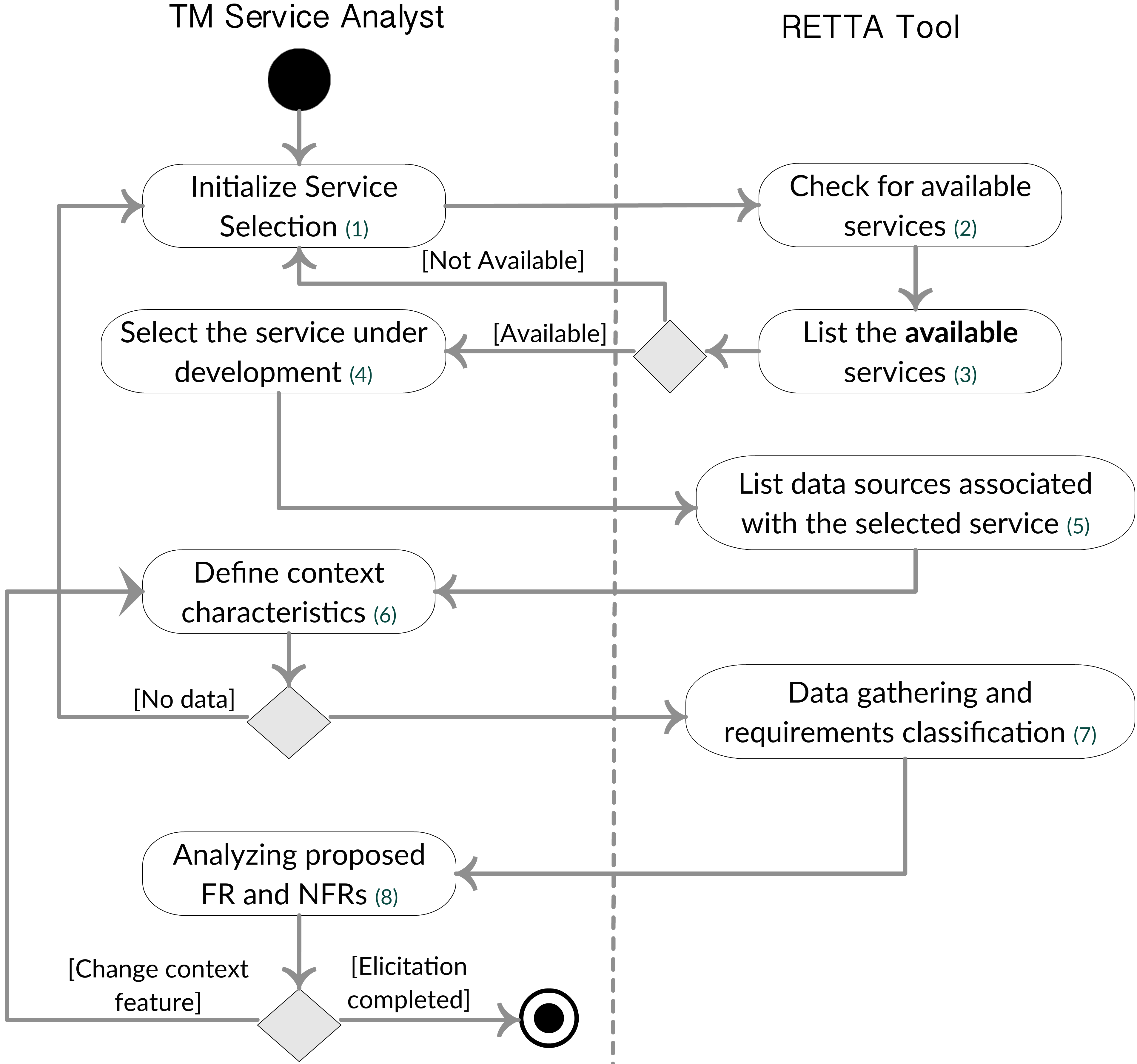}
\vspace{-2mm}
\caption{UML activity diagram of the core process of RETTA}
\label{fig:UML}
\end{figure}

The RETTA tool prototype is designed based on the lessons learned in more than five years working with transportation engineers and traffic planners. It is designed and developed using Android platform for mobile phones and tablets. Following the complexity of the computation and data analysis in the TM domain, we designed the RETTA tool in a way to be easy to develop. Thus, to design and develop the tool, we used the Model View ViewModel (MVVM) \cite{MVVM} architecture and wrapped all of the technical details and analytical computations (e.g. Shockwave model \cite{Mohammad, M2} and machine learning algorithms) to the  Model layer. Moreover, this technology reduces the coupling between the ModelView (i.e. the Controller in the Model View Controller Architecture) and the View layers, which simplifies developing tools with a high level of data analysis. Figure \ref{fig:UML} illustrates the dynamic flow and the sequence of activities of the RETTA prototype. 
\begin{figure*}
\centering

\subfloat[\scriptsize Eligible services]{\includegraphics[scale=0.1]{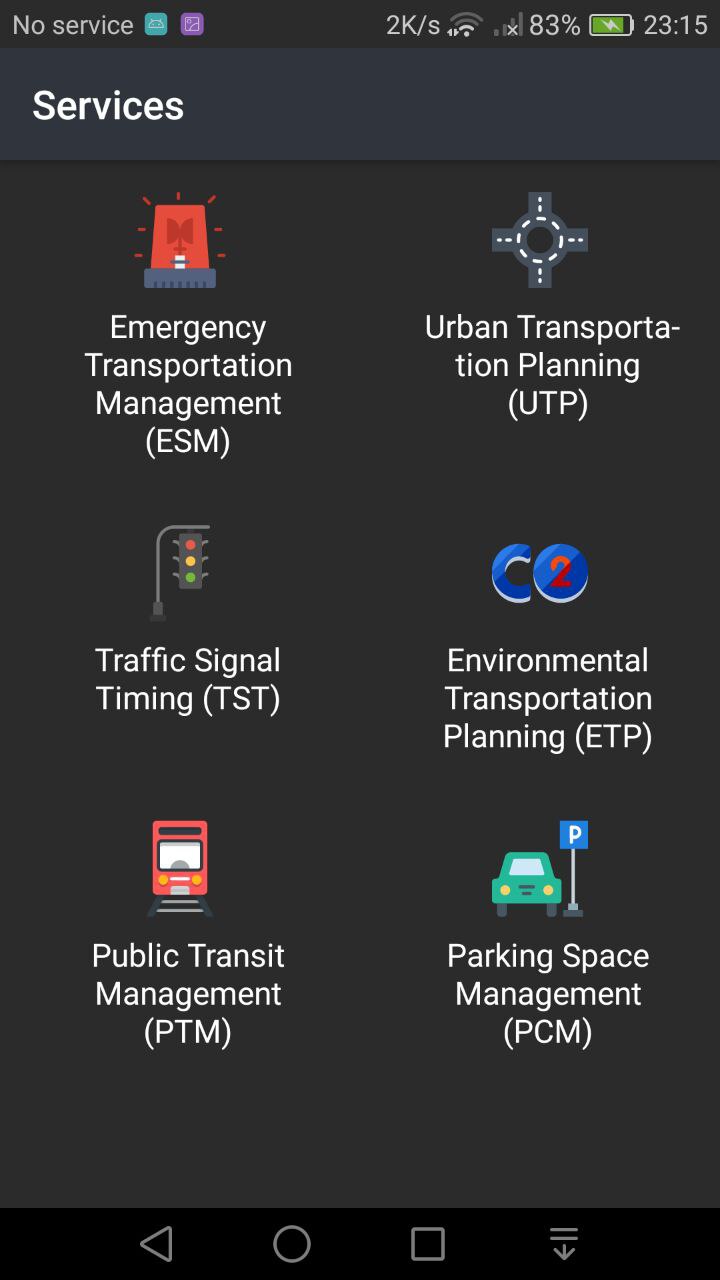}} \hspace{5mm}
\subfloat[\scriptsize TST]{\includegraphics[scale=0.1]{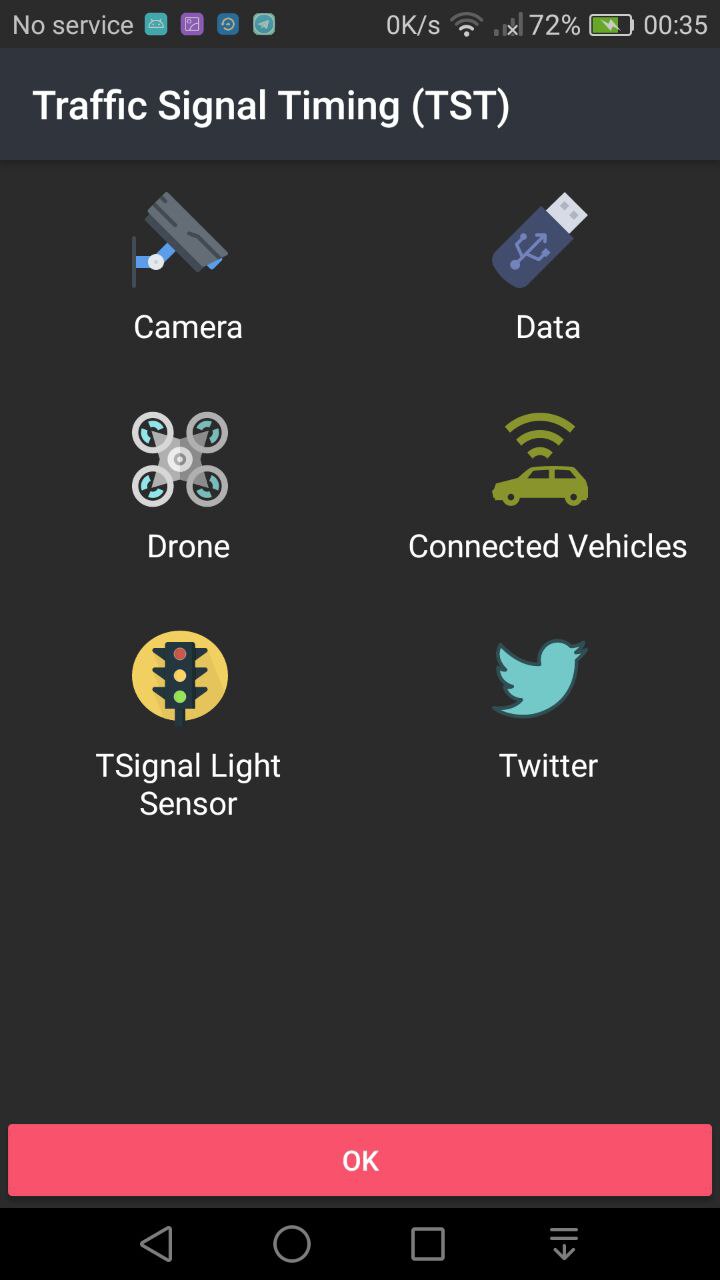}} \hspace{5mm}
\subfloat[\scriptsize Context-specific] {\includegraphics[scale=0.1]{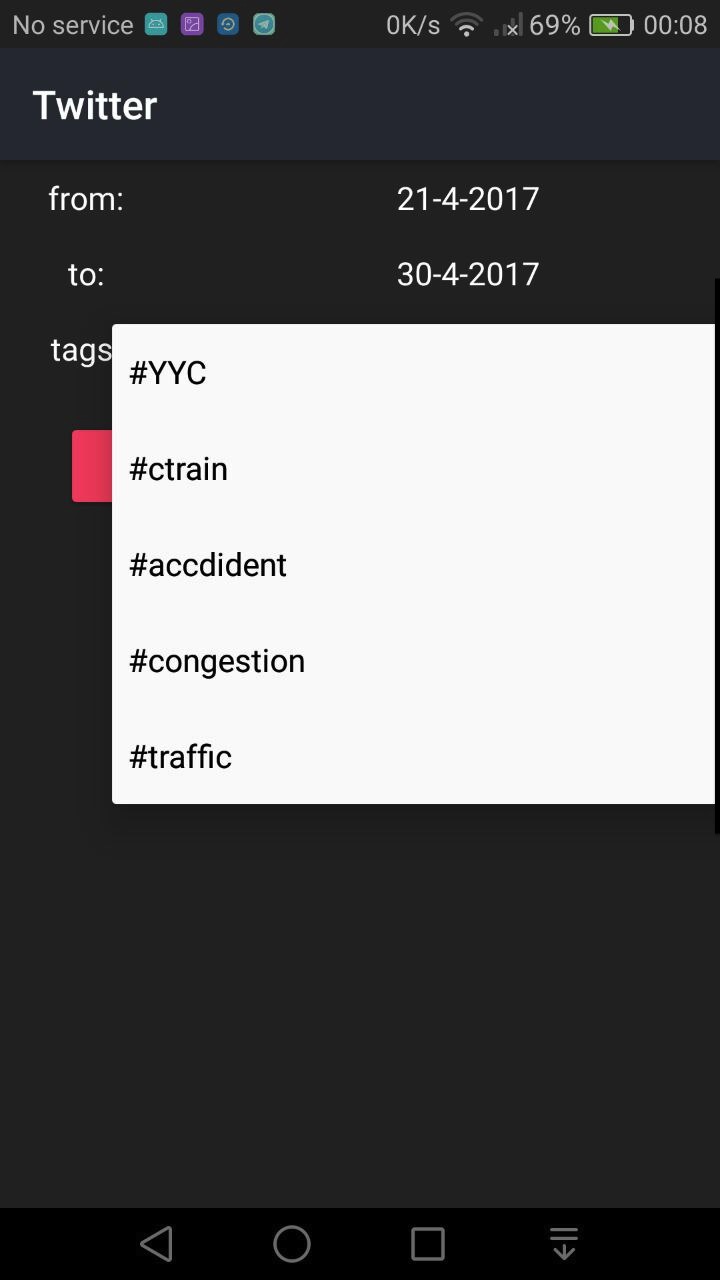}} \hspace{5mm}
\subfloat[\scriptsize FRs]{\includegraphics[scale=0.1]{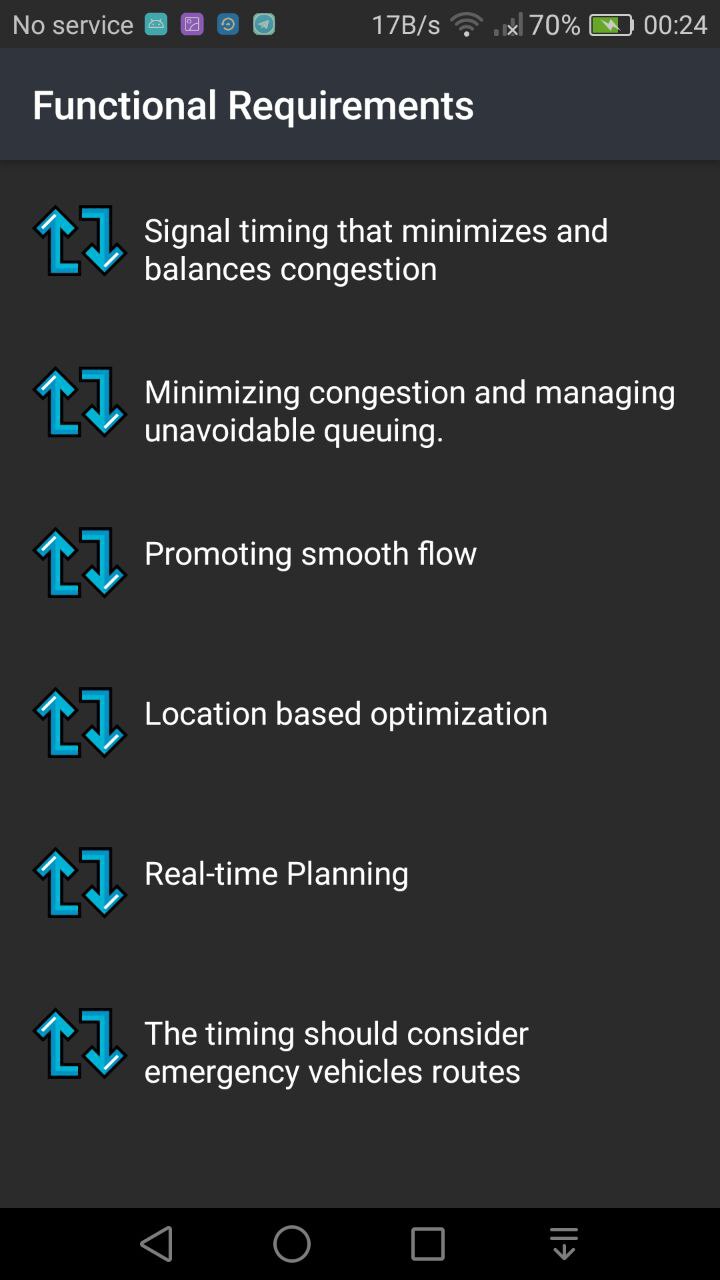}}\hspace{5mm}
\subfloat[\scriptsize NFRs]{\includegraphics[scale=0.1]{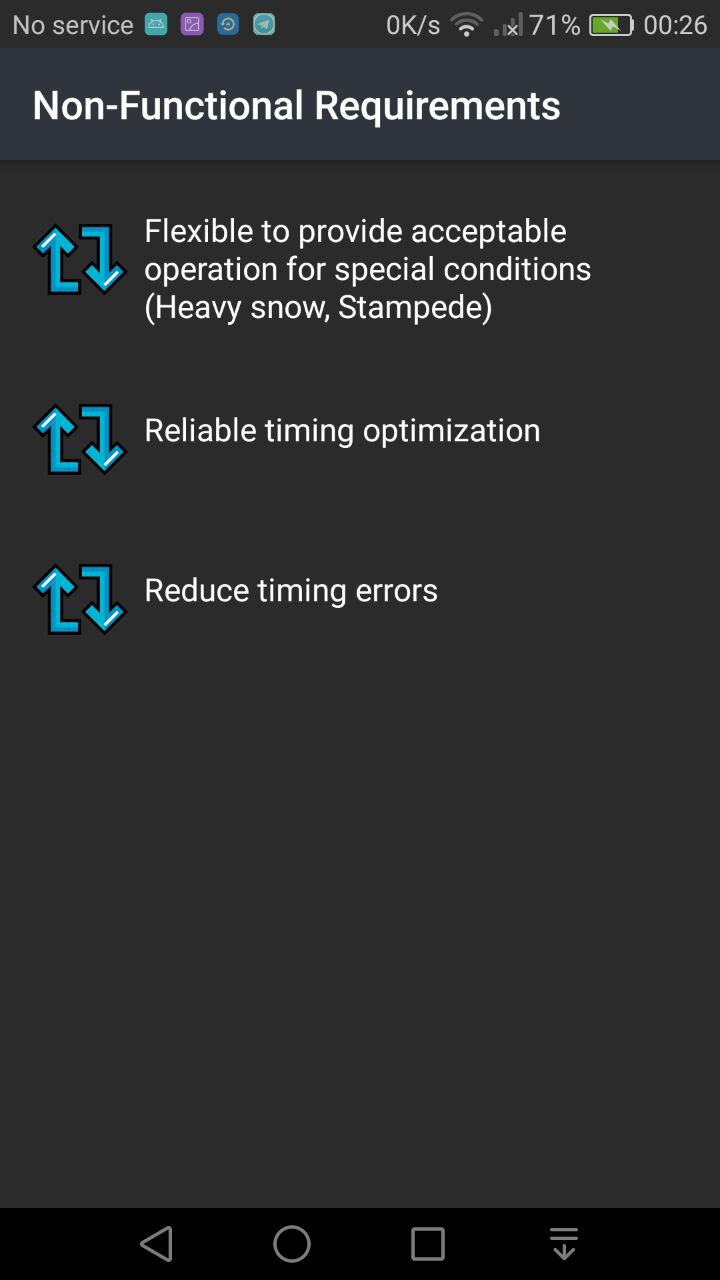}}\hspace{5mm}
\vspace{-2mm}
\caption{Detailed visual representation of classifying NFRs }
\label{fig:Retta}
\end{figure*}

Moreover, Figure \ref{fig:Retta} presents the main screens of the RETTA tool and a summarized interaction flow for identifying functional and non-functional requirements of a sample traffic management system. Once the main screen containing traffic management services such as the EMS, UTP, and TST are loaded, the user can choose the target service for which they intend to develop a software-reliant system. 
While the services listed in Figure \ref{fig:Retta} (a) all are tightly related to the TM domain, they differ in terms of the crowd that characterizes them and the data types that can be used to elicit their requirements. Figure \ref{fig:Retta} (b) shows the situation where the user aims to elicit the requirements of the TST service. Following the theoretical logics and technical parameters required for each service, there are specific data sources associated with each service. We summarize the main features of the RETTA tool by describing an {\it elicitation scenario} for {\it eliciting the TST system requirements} as follows:

TST is a TM system with the goal of optimizing signal timing in a traffic network. Once a user initializes the tool by defining the geographical location in which the TM software will be deployed  (Activity 1, Figure \ref{fig:UML}), the tool will present a list of services (Figure \ref{fig:Retta} (a)) that are qualified for implementing the requirements elicitation process  (Activities 2-3, Figure \ref{fig:UML}). In situations where the tool cannot collect adequate information for exploring the requirements of a specific service, that service will not be offered to the user. As illustrated in Figure \ref{fig:Retta} (b),  the user selects the TST service (Activity 4, Figure \ref{fig:UML}). Following this action, the RETTA tool lists available data sources which can be identified based on the geographical location of the user or the initial settings of the tool (Activities 5, Figure \ref{fig:UML}). For instance,  in an area, in which the {\it Traffic Signal  Sensor} data is not available, the tool does not list this data source. To elicit the requirements for the TST service, the following sources can be used to gather traffic data: (1) {\it real time traffic data} including the data from traffic cameras, drones, signal sensors, connected vehicles, and cellphones (e.g., via the Cities, TomTom, GoogleMap, NGSIM), (2) {\it historical traffic data} (e.g. via the Cities and transportation agencies), and (3) {\it social networks} (i.e Twitter). The requirements engineer can select any of these data sources depending on the type of the service or the context for which they are developing a software. Given the context-specific nature of TM services, in this stage, the tool asks the user to characterize the specific features of the context under development (Activity 6, Figure \ref{fig:UML}). Figure \ref{fig:Retta} (c) represents a sample screen of characterizing Twitter data collection. The details of this screen vary by the type of the data source selected by the user. For instance, the tool asks for a specific geographical area when a user selects the {\it camera} data source. Once all the required data for conducting the requirements elicitation task are collected and specified, the RETTA tool deploys various domain specific data analysis techniques (e.g. spatial-temporal traffic data analysis), and machine learning methods such as NLP, topic modelling (i.e. Latent Dirichlet
Allocation (LDA)), association rule mining and Na\"{i}ve Bayes algorithms to elicit and classify TM services' requirements (Activity 7, Figure \ref{fig:UML} and Figure \ref{fig:Retta} (d-e)). Given that due to the sparsity of word co-occurrence patterns in short texts such as Twitter and microblogs \cite{LDA}, the RETTA tool combines all of the retrieved tweets related to a specific service and then applies the LDA algorithm. The main data pre-processing steps such as removing English stop-words, numbers, HTML tags, and stemming process will be applied before applying the topic modeling and NLP approaches on the retrieved tweets. Moreover, to classify the requirements, we defined some domain specific {\it regular expressions} \cite{DataTrack} and increased the weighting of valuable words for each TM service. For instance, {\it malfunction, signal, light, traffic, accident} are the keywords that can characterize the reliability of TST service. 

%{\it ``Traffic really slow eastbound {\#hwy1} headed into \#YYC \#yyctraffic  wonder if there's an accident?''} and {\it ``Another ACCIDENT at 1A St \& 53rd Ave SW! Traffic control device needed before someone is killed! @cityofcalgary @gccarra @CalgaryPolice \#YYC''}, {\it ``Lots of light signal malfunctions in NW \#yyc. \#yyctraffic''}

\section{Conclusion and Future Work}
The RETTA tool provides an interactive environment for eliciting requirements in the traffic management domain. A short informal evaluation of the tool has been carried out in Intelligent Software Systems Laboratory at the University of Calgary. Software developers found the RETTA remarkably easy to use and very thought provoking for eliciting traffic domain requirements. Future work will concentrate on improving the efficiency and completing the text analysis and classification approaches for exploring and classifying requirements.  Moreover, we aim to improve the applied requirements elicitation process in the RETTA tool to address the complexity and the scale of the crowd and to ensure that we record their requirements efficiently and precisely.
\vspace{-1mm}

\bibliographystyle{IEEEtran}
\bibliography{IEEEabrv,References}
\end{document}